-------------------------------------------------------------------------
\documentstyle[12pt]{article}
\def \lsim { ^< \hspace*{-7pt} _\sim}
\def \gsim { ^> \hspace*{-7pt} _\sim}

\newcommand{\be}[1]{\begin{equation} \label{(#1)}}
\newcommand{\ee}{\end{equation}}
\newcommand{\ba}[1]{\begin{eqnarray} \label{(#1)}}
\newcommand{\ea}{\end{eqnarray}}

\newcommand{\rf}[1]{~(\ref{(#1)})}

\begin{document}

\begin{center}
{\bf
   On leading charmed meson production in $\pi$-nucleon interactions
}
\bigskip

     V.A.~Bednyakov       \\[1mm]
{\it
     Joint Institute for Nuclear Research, Dubna, Russia    \\
   }
\end{center}

\begin{abstract}
It is shown that the $D$--meson,
whose light quark is the initial-pion valence quark
and whose charmed quark is produced in annihilation of valence
quarks and has got a large enough momentum, is really a leading meson
in reactions like $\pi^- p \rightarrow DX$.
If such annihilation of valence quarks from initial hadrons
is impossible there must be no distinct leading effect.
\end{abstract}

\vspace*{0.5cm}

Recently the E769 collaboration \cite{E769} has reported confirmation
of previously obtained \cite{NA32} enchanced leading production
of $D^\pm$- and $D^{*\pm}$-mesons in 250 GeV $\pi^\pm$--nucleon
interaction.
A leading charmed meson is considered to be one with the longitudinal
momentum fraction $x_F^{}>0$, whose light quark (or anti-quark) is
of the same type as one of the quarks in the beam particle.
At large $x^{}_F$ significant asymmetry was found:
\be{Asymmetry}
A(x^{}_F) \equiv \frac{\sigma(\mbox{leading})-\sigma(\mbox{non-leading})}
                      {\sigma(\mbox{leading})+\sigma(\mbox{non-leading})}.
\ee

Such asymmetry for the production of charmed hadrons is not
expected in perturbative quantum chromodynamics.

Some years ago
a simple non-perturbative mechanism of leading charmed mesons production
was considered \cite{Bedny}
for data analysis of CERN experiment on $D$-mesons
production in $\pi^-p$-collisions \cite{Aguil}.
It was demonstrated that
presence of a valence quark from the initial pion
(so--called leading quark state) in the final charmed meson is a
necessary but insufficient condition for the meson to be a leading one.
Actually, those $D$ are leading mesons whose light quarks are
valence quarks of the pion {\em and\/} charmed quarks are produced in
annihilation of valence quarks and carry a large momentum $x_c$.
\bigskip

The leading effect is a characteristic property of inclusive
production of charmed hadrons \cite{Basile}.
A hadron $H$\ produced in the reaction $a+b \rightarrow H + \ldots$
and carrying the largest portion of the momentum, $p^{}_H=O(\sqrt{s}/2)$,
is regarded as a leading hadron.
The corresponding momentum spectrum $dN/dx^{}_F$
usually parametrised in the form $(1- x_F^{})^n$ at a large Feynman
variable $x_F^{}= \frac{2}{\sqrt{s}}P_\|$ is ``hard'' for leading hadrons
$(0 < n~ \lsim ~3)$ and ``soft'' for non--leading ones $(n~ \gsim ~5)$.

In the quark--parton approach the leading charmed meson $H$\
is a result of recombination of the spectator valence quark $q_v^{}$
with the charmed quark produced in a parton subprocess.
Owing to the large momentum of the valence quark $x_v^{}$
$H$\ turns to be a leading meson,
its momentum is large enough $x_H^{} = x_v^{} + x_c^{} > x_v^{}$.

{}From this point of view
$D^-(d\bar{c})$ and $D^\circ(\bar{u}c)$
directly produced
in the reaction $\pi^-(d\bar{u})+ p\rightarrow D(d\bar{c};\bar{u}c) + X$
must be both leading mesons, i.e. yields of
$D^-(d\bar{c})$ and $D^\circ(\bar{u}c)$ have to be practically the same
at large momentum (say, $x^{}_F {} > 0.5$).

On the other hand, let us assume for a moment
that hadrons consist of valence quarks alone.
This picture takes place,
for instance, in deep inelastic phenomena
at quite large $x_F^{}$, when all non-singlet
parton distribution functions vanish.

In this case $D^\circ(\bar{u}c)$--mesons can by no means
result form the reaction $\pi^-(d\bar{u})+p(uud)\rightarrow D+X$
because there is no parton subprocess which can ensure
$c$-quark creation.
On the other hand, the $\bar{c}$--quark appears
due to valence quarks annihilation
$\bar{u}^\pi_v u^p_v \rightarrow c\bar{c}$,
providing the $D^-(d\bar{c})$--meson in the final state.
It is clear 
that some difference in $\pi^-$-nucleon production of
leading $D^\circ(\bar{u}c)$ and $D^-(d\bar{c})$--meson
has to take place at sufficiently large $x_F^{}$.
To demonstrate this feature quantitatively
let us follow briefly the work \cite{Bedny}.

The invariant differential cross section for the process
$\pi^- p \rightarrow D\, X$ in the centre--of--mass system
at the energy $\sqrt{s}$ and $x_F^{}>0$ can be written down
in the form \cite{Likhoded}:
\be{CS}
x^*\frac{d\sigma}{dx\, dp^2_T} =
        \exp{\{-2p^2_T/\sqrt{s}\}} \int R(x_{sp},x_c;x)
        \frac{dx_{sp}}{x_{sp}} \frac{dx_c}{x^*_c}
      \biggl\{\frac{x^*_c x_{sp} d\sigma}{dx_{sp} dx_c dp^2_T} \biggr\}.
\ee
        Here $x\equiv x_F^{}$, $x_{sp}$, $x_c$ are the
        Feynman variables of $D^-(D^\circ)$--meson, spectator
        $d(\bar{u})$- and produced $\bar{c}(c)$--quark;
        $x^*_{} = 2E_D/\sqrt{s}$, $x^*_c= 2E_c/\sqrt{s}$.

        The phenomenological
        recombination function \cite{Likhoded}, \cite{Takasugi}
        $R(x_{sp},x_c; x) \sim  
                          \delta(x - x_{sp}- x_c)$
        provides a probability of producing a $D^-(D^\circ)$--meson
        (with the momentum $x$) by means of a $d(\bar{u})$--quark
        $(x_{sp})$ and a $\bar{c}(c)$--quark $(x_c)$.

        The probability of existence of spectator $d(\bar{u})$-quark and
        charmed $\bar{c}(c)$--quark is determined by the expression:
\be{cs}
   \frac{x^*_c x^{}_{sp}\,d\sigma}{dx_{sp} dx_c dp^2_T}
        = x_{sp} \int dx_L dx_R
\sum_{i = q,\bar{q},g} f^\pi_{d(\bar{u}) i}(x _{sp},x_L) f^p_{\bar{i}}(x_R)
\frac{x^*_c d\sigma}{dx^{}_c dp^2_T}.
\ee
        Here $\frac{x^*_c d\sigma}{dx^{}_c dp^2_T}$ is
        the quantum--chromodynamics cross section
        for the charm production parton subprocess
        $i\bar{i} \rightarrow c\bar{c}$ \cite{Gluck}.
        The single-particle proton distribution functions,
        $f^p_{i}(x_R^{})$, are extracted from deep inelastic
        lepton-proton scattering \cite{Zlat}.
        The analytical form of two-particle pion distribution functions,
        $f^\pi_{vi}(x_{sp},x_L)$,
        is given in the statistical parton
        model \cite{Likhoded}, \cite{MPDF}.
        The free parameters of these analytical forms
        can be fixed via comparison with the data. 

        It is clear from relation\rf{cs}
        that the above-mentioned difference in yields of
        $D^\circ(\bar{u}c)$ and $D^-(d\bar{c})$-- mesons
        mainly arises due to different contributions of
        distribution functions:
        $\sum f^\pi_{vi}\cdot f^p_{\bar{i}}$.

        For a $D^\circ$--meson the sum is
\be{D0}
   \sum {D^\circ}= f^\pi_{vv} \cdot f^p_s + f^\pi_{vs}\cdot
                (3f^p_v + 6f^p_s).
\ee
        For a $D^-$--meson we have
\be{Dminus}
  \sum {D^-} = f^\pi_{vv} \cdot f^p_s + f^\pi_{vs}\cdot
         (3f^p_v + 6f^p_s) +
        2 f^\pi_{vv} \cdot f^p_v =
        \sum {D^\circ} + 2f^\pi_{vv} \cdot f^p_v,
\ee
        where index $v$ corresponds to valence quarks and $s$
        to sea quark.
        For simplicity flavour symmetric distributions were used and
        the gluon contribution was omitted.

        Therefore the total momentum
        spectrum of $D^-$ and $D^\circ$--meson production
        in $\pi^-p$-collisions can be put down in the form
\be{totalsp}
        \frac{d\sigma}{dx}(D^- + D^\circ) =
        2 \frac{d\sigma}{dx}(D^\circ)+\frac{d\sigma}{dx}(v).
\ee

        This formula was used for fixing 
        distribution functions $f^\pi_{vi}$\
        by means of comparison with the data on leading
        $D$--meson production in $\pi^-p$--collisions
        at $\sqrt{s}=26$~GeV \cite{Aguil}.

        It was obtained that the "valence" component,
        $\frac{d\sigma}{dx}(v)$,
        due to "hard" shape of valence distributions,
        ensured the non-vanishing
        total spectrum for $x^{}_F {~}\gsim ~0.5$.
        At low $x_F^{}$ the total spectrum was saturated by the other
        component -- $\frac{d\sigma}{dx}(D^\circ)$.

        The term $\frac{d\sigma}{dx}(v)$
        makes no contribution to the spectrum of $D^\circ$--mesons
        (see formula\rf{D0}), therefore
        the yield of neutral $D^\circ$--mesons
        at large $x_F^{}$ is small enough.

        Figure 1 shows the ratio:
\be{Ratio}
        R(x_F^{})= \frac{\frac{d\sigma}{dx}(\pi^-p \rightarrow D^\circ X)}
     {\frac{d\sigma}{dx}(\pi^- p \rightarrow D^-X)},
\ee
        which quantitatively illustrates the suppression of
        the $D^\circ$ yield as comparied with the $D^-$ one.
        The experimental points are
        recalculated from combined data on
        asymmetry $A$\rf{Asymmetry} measured on nuclei \cite{E769}.
        The curves obtained in paper \cite{Bedny}
        and considered as a predictions  successfully
        fit the new data \cite{E769}.

        Figure 2 shows two curves for asymmetry $A$\rf{Asymmetry},
        calculated on the basis of the ratio \rf{Ratio}.
        The curves also describe the data well.

\bigskip

        Thus it is demonstrated that presence of a valence quark
        from the initial hadron (as a spectator)
        in the final charmed meson is a necessary
        but insufficient condition for the meson
        to have a "hard" momentum spectrum (i.e. to be a leading meson).

        Actually, the $D$-meson is a "real" leading meson
        whose light quark is a spectator valence quark and charmed quark
        (anti-quark) is produced in annihilation of valence quarks
        from initial hadrons.

\bigskip

        In addition, it is easy to
        construct relations like\rf{Ratio} for reactions similar to
        $\pi^- p \rightarrow DX$. Thus we have for
        $x_F > 0.5$ (denominators show the leading mesons):
$$
\frac{\sigma (\pi^+ n \rightarrow D^+ X)}
     {\sigma (\pi^+ n \rightarrow \bar{D}^\circ X)} =
\frac{\sigma (\pi^+ \bar{p} \rightarrow \bar{D}^\circ X)}
     {\sigma (\pi^+ \bar{p} \rightarrow D^+ X)} =
\frac{\sigma (\pi^- \bar{n} \rightarrow D^- X)}
     {\sigma (\pi^- \bar{n} \rightarrow D^\circ X)} = R(x_F^{});
$$
$$
\frac{\sigma (K^- p \rightarrow \bar{D}^\circ X)}
     {\sigma (K^- p \rightarrow D^-_s X)} =
\frac{\sigma (K^+ \bar{p} \rightarrow D^\circ X)}
     {\sigma (K^+ \bar{p} \rightarrow D^+_s X)} = R(x_F^{});
$$
$$
\frac{\sigma (\pi^- \bar{p} \rightarrow D^- X)}
     {\sigma (\pi^- \bar{p} \rightarrow D^\circ X)} =
\frac{\sigma (\pi^+ p \rightarrow D^+ X)}
     {\sigma (\pi^+ p \rightarrow \bar{D}^\circ X)}=
\frac{\sigma (\pi^- n \rightarrow D^\circ X)}
     {\sigma (\pi^- n \rightarrow D^- X)} = 2 R(x_F^{});
$$
$$
\frac{\sigma (\pi^+ \bar{n} \rightarrow \bar{D}^\circ X)}
     {\sigma (\pi^+ \bar{n} \rightarrow D^+ X)}= 
\frac{\sigma (K^- n \rightarrow \bar{D}^\circ X)}
     {\sigma (K^- n \rightarrow D^-_s X)} =
\frac{\sigma (K^+ \bar{n} \rightarrow \bar{D}^\circ X)}
     {\sigma (K^+ \bar{n} \rightarrow D^+_s X)} = 2 R (x_F^{}).
$$

\vspace*{1cm}

{\bf Figure Captions      }

Fig.~1. $D^\circ$--to--$D^-$ yield ratios\rf{Ratio} for
        $\pi^- p$--collisions (lower curve) and
        $\pi^- n$--collisions (upper curve).
        The points are recalculated from the data on
        asymmetry $A$ \cite{E769}.

Fig.~2.
        Asymmetry $A$\rf{Asymmetry}
        on the proton target (upper curve)
        and the neutron target (lower curve)
        calculated on the basis of the ratio\rf{Ratio}.
        The data from ref. \cite{E769}.

\end{document}